\documentstyle[11pt,aaspp4]{article}

\begin{document}

\def\Rm{R_m }

\def\dcad{{\eta}_{AD} }
\def\dcd{{\eta}_{d} }
\def\dco{{\eta}_{Ohm} }

\def\hal{H$\alpha$ }
\def\sodi{{Na}$^{+}$ }
\def\hyd{H$_2$ }

\def\kms{km s$^{-1}$\ }
\def\cms{cm s$^{-1}$\ }
\def\cmc{cm$^{-3}$\ }
\def\cmss{cm$^{2}$ s$^{-1}$\ }
\def\cmcs{cm$^{3}$ s$^{-1}$\ }

\def\msun{M$_\odot$\ }
\def\mj{M$_J$\ }
\def\Lhal{L_{H\alpha} }
\def\Lx{L_X }
\def\Lbol{L_{bol} }
\def\Fhal{F_{H\alpha} }

\def\teff{T$_{e\! f\! f}$~}
\def\vsini{{\it v}~sin{\it i}~}

\title{Rotation \& Activity in Mid-M to L Dwarfs}

\author{Subhanjoy Mohanty\altaffilmark{1}, Gibor Basri\altaffilmark{1}}

\altaffiltext{1}{Astronomy Department, University of California, Berkeley}

%\index{Rotation Velocity}
%\index{Chromospheric Activity}
%\index{M Dwarfs}
%\index{L Dwarfs}

\begin{abstract}
We analyze rotation velocities and chromospheric (\hal) activity, derived from multi-year, high-resolution spectra, in 56 mid-M to L dwarfs.  Rotation velocities are found to increase from mid-M to L.  This is consistent with a lengthening of spin-down timescale with later type, though in the L types the trend may also be a function of stellar age.  From M5 to M8.5, a saturation-type rotation-activity relation is seen, similar to that in earlier types.  However, the saturation velocity in our case is much higher, at $\sim$ 12 \kms.  A sharp drop in activity is observed at $\sim$ M9, with later types showing little or no \hal emission, in spite of rapid rotation.  This may be due to the very high resistivities in the predominantly neutral atmospheres of these cool objects.      
\end{abstract}

\section{Introduction}
In early M dwarfs (M1 - M5), the trend of lengthening spindown timescale with later type, observed in G and K dwarfs, continues: early M's are faster rotators than coeval G and K stars, and there is some evidence for increasing \vsini from M1 to M5 (\cite{delfosse98}).  Two mechanisms have been proposed to explain this trend:  the deepening of the convective envelope with later type (for stars with M$>$0.3\msun, i.e., earlier than $\sim$ M2.5 on the ZAMS; see \cite{charb97}, and the small-scale nature of the turbulent magnetic field (for less massive, fully convective stars; see \cite{durney93}).

The rotation-activity relation in the M1-M5 dwarfs is also similar to that in G and K stars.  A saturation-type relation is seen in coronal and chromospheric activity indicators (\cite{delfosse98}).  Above a critical rotation velocity ($\sim$ 5\kms for coronal X-ray and $\sim$ 2\kms for chromospheric \hal emission), activity in the early M's is saturated (at $\sim$ 10$^{-3}$ for  $\Lx/\Lbol$ and $\sim$ 10$^{-3.5}$-10$^{-4}$ for $\Lhal/Lbol$).  Below this velocity, a range of activity levels, less than the saturation limit, are seen (though the observational sensitivity in \vsini is too low to probe the exact trend of activity with rotation in this regime) (Fig. 1).  In stars earlier than $\sim$ M2.5, these data are consistent with the standard rotation-activity paradigm.  In this picture, the magnetic fields that drive activity are generated by an $\alpha\Omega$ dynamo, whose efficiency strongly increases with faster rotation (\cite{charb97}).  Faster rotation thus leads to more activity; saturation sets in when the fields are strong enough to cover the entire stellar surface.  One might expect the rotation-activity relation to weaken at about M2.5, since  later spectral types are fully convective.  As such, they may have only a turbulent dynamo, which is only marginally rotation-dependent (\cite{durney93}).  However, no break in the rotation-activity relation is observed at $\sim$ M2.5; indeed, it holds all the way down to $\sim$ M5 (\cite{delfosse98}).  Perhaps the turbulent dynamo becomes dominant even {\it before} full convection sets in, and is efficient enough, even though only weakly rotation-dependent, to induce saturation at low \vsini in these stars.  This would induce a smooth transition in the rotation-activity relation across the full-convection boundary.  
 
In this paper, we study the \vsini and chromospheric \hal emission in 56 mid-M to L dwarfs (M5 - L6), all of which are field objects.  Our aim is to examine whether the trends in the rotation and rotation-activity connection noted above continue as we move beyond early M, to still later spectral types.  A preliminary version of this work was presented at Cool Stars XI by Basri (2000).  In the present work, the previous sample has been extended to include more objects, and \vsini, \hal flux, spectral type and \teff are determined more accurately.

\section{Rotation Velocity}
In Fig. 2 we show the \vsini derived for our sample.  It is immediately obvious that rotational velocity continues to increase with later spectral type.  Most stars in the range M5-M8 in our sample are slow rotators (\vsini $<$ 15 \kms), about half the stars from M8.5-L1 are fast rotators ($>$ 15 \kms), and {\it all} the objects L2 and later are very rapid rotators ($>$ 20 \kms).  This last fact is remarkable:  unlike in the earlier types, where some slow rotators are always present, we see no slow rotators at types L2 and later, even though \vsini is only a {\it lower} limit on the rotational velocity (so the real velocity is even higher).  

Most of our M dwarfs show no signs of Lithium, so they are presumably stellar, with ages $>$ 1 Gyr.  In the \cite{delfosse98} sample, all stars earlier than $\sim$ M5, and with kinematic ages consistent with the young disk ($\sim$ 3 Gyr), have \vsini $<$ 10 \kms.  In their old disk ($\sim$ 10 Gyr) sample, all stars earlier than M5 have \vsini $<$ 5 \kms.  Therefore, even if our stellar M dwarf sample (i.e., those with no Lithium) were composed only of young disk objects, it would still be true that the mid-M dwarfs rotate at least as fast as, and the late M dwarfs rotate much faster than, earlier M dwarfs of roughly the same age.  This conclusion is only strengthened if our sample comprises old disk or halo objects.  In other words, the trend of increasing spindown timescale with later type continues unabated down to late M.  In the L dwarfs, however, the question of age is more complicated.  There is an observational bias towards finding young L types.  Objects later than about L2 are expected to be brown dwarfs.  With time, these cool down, moving down the L spectral sequence and becoming fainter and harder to detect.  Below $\sim$ L2, therefore, a magnitude limited survey selects for comparatively younger objects.  For example, those L dwarfs in our sample shown to be substellar by the Lithium test are all less than 1 Gyr old, or well within the young disk population.  Those that do not show Lithium are more massive and comparatively older, but could still be very young.  Hence it is not clear whether the high \vsini we observe in the L dwarfs is a true signature of increasing spindown timescale, or simply a result of their being young objects (and thus not having had time to spindown).  This question will have to be sorted out with a larger sample of L dwarfs, and better age determinations, both through the Lithium test and kinematics.  

\section{Rotation-Activity Relation}
In Fig. 3 we plot chromospheric activity, as measured through \hal emission, against \vsini for our sample.  Activity levels are denoted in two ways:  by the ratio of \hal luminosity to bolometric luminosity ($\Lhal/\Lbol$), and by the \hal surface flux ($\Fhal$) (there is no consensus at present on whether $\Lhal/\Lbol$ or $\Fhal$ is physically the more fundamental measure of activity).  The top panels show only M5-M8.5 dwarfs, the bottom ones show our entire sample (M5-L6).  Two conclusions may immediately be drawn from the plots.

First, M5-M8.5 dwarfs evince a rotation-activity relation very similar to that in earlier M's.  A saturation-type relation is seen:  above a critical velocity, activity is saturated (at $\Lhal/\Lbol$ = 10$^{-3.5}$-10$^{-4}$, precisely as observed in earlier M's), while below the critical velocity, a range of unsaturated activity levels exist.  In other words, the rotation-activity connection observed in the early M dwarfs continues down to late M.  The only significant change is in the magnitude of the critical velocity.  In our sample it is $\sim$ 12 \kms, while for M1 - M5 dwarfs it is about 2 \kms.  Moreover, we find no significant correlation between rotation and activity below 12 \kms (binning the M5-M8.5 dwarfs into smaller spectral groups (see Fig. 3) does not change this).  This is consistent with the operation of a turbulent dynamo that is only marginally dependent on rotation.  The suggestion is that below \vsini $\sim$ 12 \kms, the dynamo is comparatively inefficient and effectively independent of rotation, leading to a spectrum of unsaturated activity levels.  Above 12 \kms, however, the rotation rate is finally high enough to affect the dynamo efficiency, eventually leading to saturation.  In that case, why saturation occurs at a much lower \vsini of 2 \kms in M2.5-M5 dwarfs (which are also fully convective, and so presumably support a turbulent dynamo as well) is a question that needs to be resolved.  

Second, the rotation-activity connection appears to break down around M9.  Among the L types, the highest levels of $\Lhal/\Lbol$ we observe are at least an order of magnitude lower than saturated levels in stars of type M8.5 and earlier.  Most of the L dwarfs have only upper limits in \hal emission; true activity levels in these may be much lower.  These results are surprising given that, on average, L dwarfs rotate much faster than the M's.  Our observations are not biased by any difficulty in detecting \hal in the L's:  at the low \teff in these objects, chromospheric \hal is even easier to detect against the fainter photosphere than in the hotter M types.  If the absolute levels of \hal emission were the same in the L's and M's, $\Lhal/Lbol$ would actually be greater in the L dwarfs.  Instead, as Fig. 3 shows, the \hal surface flux itself declines in the L's, and precipitiously enough for even $\Lhal/\Lbol$ to decrease.  We note that M9-M9.5 dwarfs appear to bridge the gap in activity levels between the earlier M dwarfs and the L types.    

Basri (2000) suggested that very rapid rotation may damp turbulence, leading to an inefficient turbulent dynamo and hence low levels of magnetically driven activity, in late M and L dwarfs.  This now seems unlikely, given that the fastest rotator (Kelu-1:  L2, 60 \kms) actually produces one of the highest \hal emissions among these objects.  A more probable solution (\cite{hof99}, \cite{basri00}) is that the falloff in activity with later type is actually a consequence of decreasing \teff (see Fig. 4).  The resistivities in the cool, mostly neutral atmospheres of late M and L dwarfs are likely very high.  Consequently, electric currents may be rapidly damped, preventing the build-up of large magnetic stresses in the atmospheres of these objects.  Since, in the conventional picture, the eventual dissipation of such stresses energetically supports a corona and chromosphere and drives activity, \hal emission would be severely depressed in late M and L dwarfs.  This view is supported by recent calculations of resistivity in late M and L atmospheres by \cite{mohantysub}.  

Finally, we note that high resistivities effectively decouple the magnetic field from the atmosphere, causing inefficient angular momentum loss through magnetically driven winds.  This may contribute to the high \vsini observed in the L dwarfs.  

\section{Conclusions}
\begin{itemize}
\item Rotation velocities continue to increase from mid-M to L.  In the M dwarfs, this continues the trend of lengthening spindown timescale with later type, that is seen in earlier types.  In the L dwarfs, this trend may be due to substellarity and thus comparative youth, or a true increase in spindown timescale, perhaps due to the high atmospheric resistivities.
\item A saturation-type rotation-activity connection is observed from M5-M8.5, similar to that in the early M's.  However, the critical velocity for saturation ($\sim$ 12 \kms) is much higher than in the early M's ($\sim$ 2 \kms).  No obvious correlation is seen between rotation and activity below 12 \kms.  This, together with the high critical velocity, may be evidence for a turbulent dynamo that is only weakly dependent on rotation.
\item A sharp drop in activity is seen at $\sim$ M9, with later types showing little or no evidence of \hal emission, in spite of being very rapid rotators.  This may be due to the very high resistivities in the cool, predominantly neutral atmospheres of late M and L dwarfs:  the resultant rapid decay of currents damps the build-up of large magnetic stresses, which could otherwise energetically support a chromosphere and corona and drive activity.  
\end{itemize}

\plotfiddle{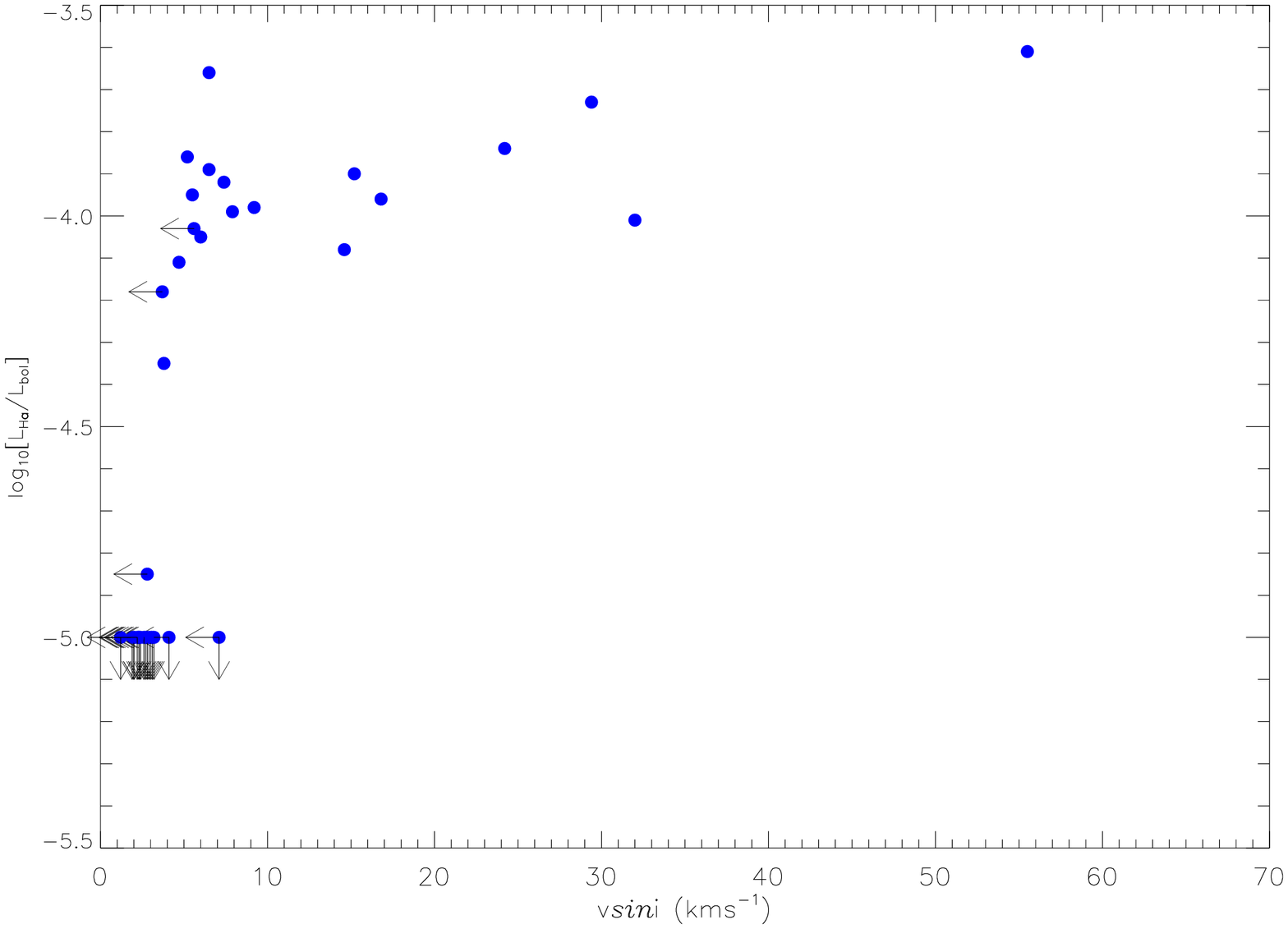}{8cm}{0}{45.}{45.}{-150cm}{0cm}
\figcaption{\label{fig1} $\Lhal/\Lbol$ versus \vsini for M4 and M5 dwarfs (from \cite{delfosse98}).  Upper limits in \hal emission and \vsini marked by arrows (though actual equatorial velocity may be higher) }

\plotfiddle{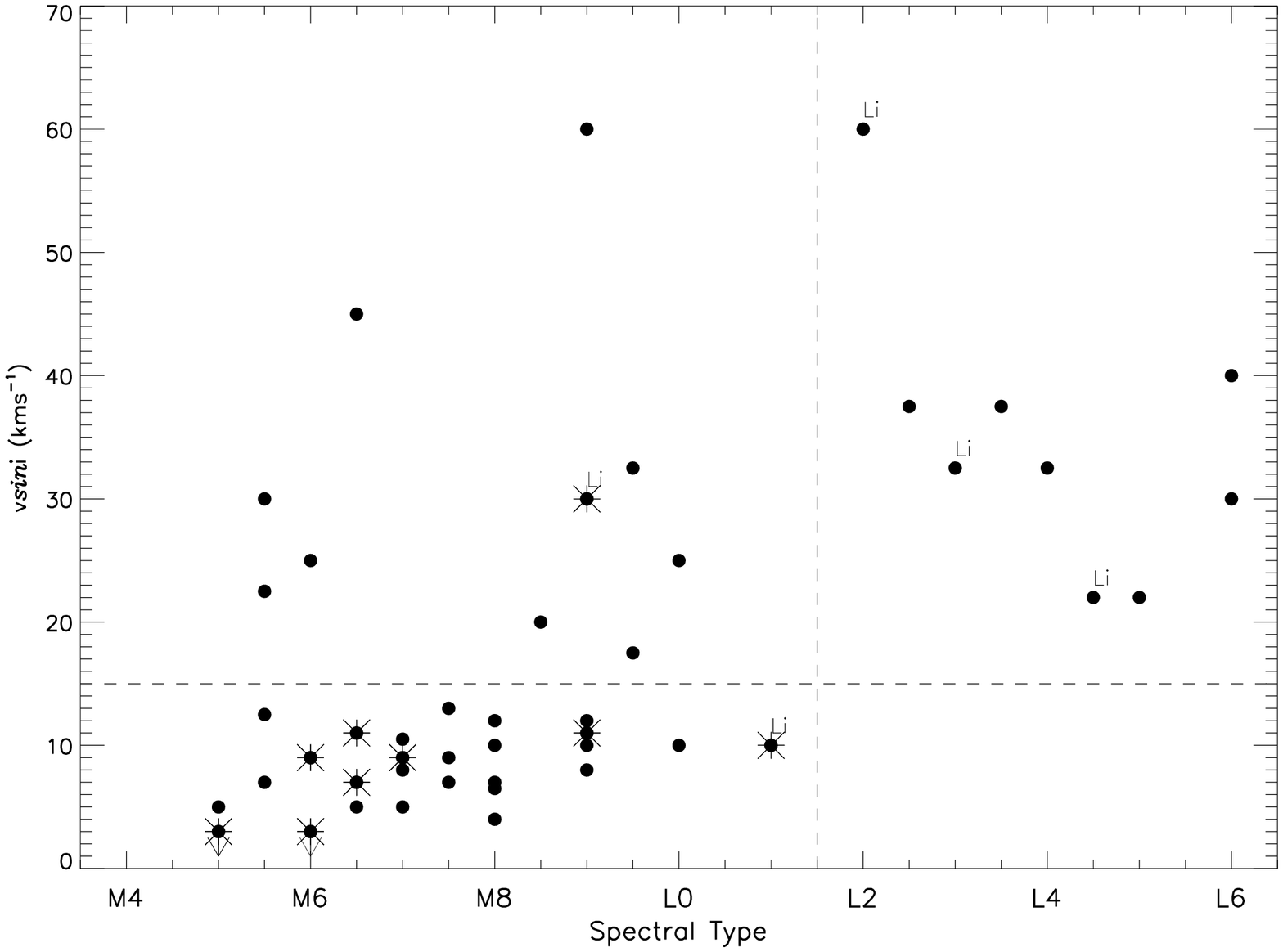}{7cm}{0}{45.}{45.}{-150cm}{0cm}
\figcaption{\label{fig2} \vsini versus Spectral Type, for M5 to L6 dwarfs.  Overlapping objects marked with spines.  Horizontal line is at 15 \kms, below which we define objects as slow rotators.  Vertical line marks spectral type L1.5, above which all objects rotate rapidly.  `Li' marks objects with Lithium; these are confirmed brown dwarfs. }

\plottwo{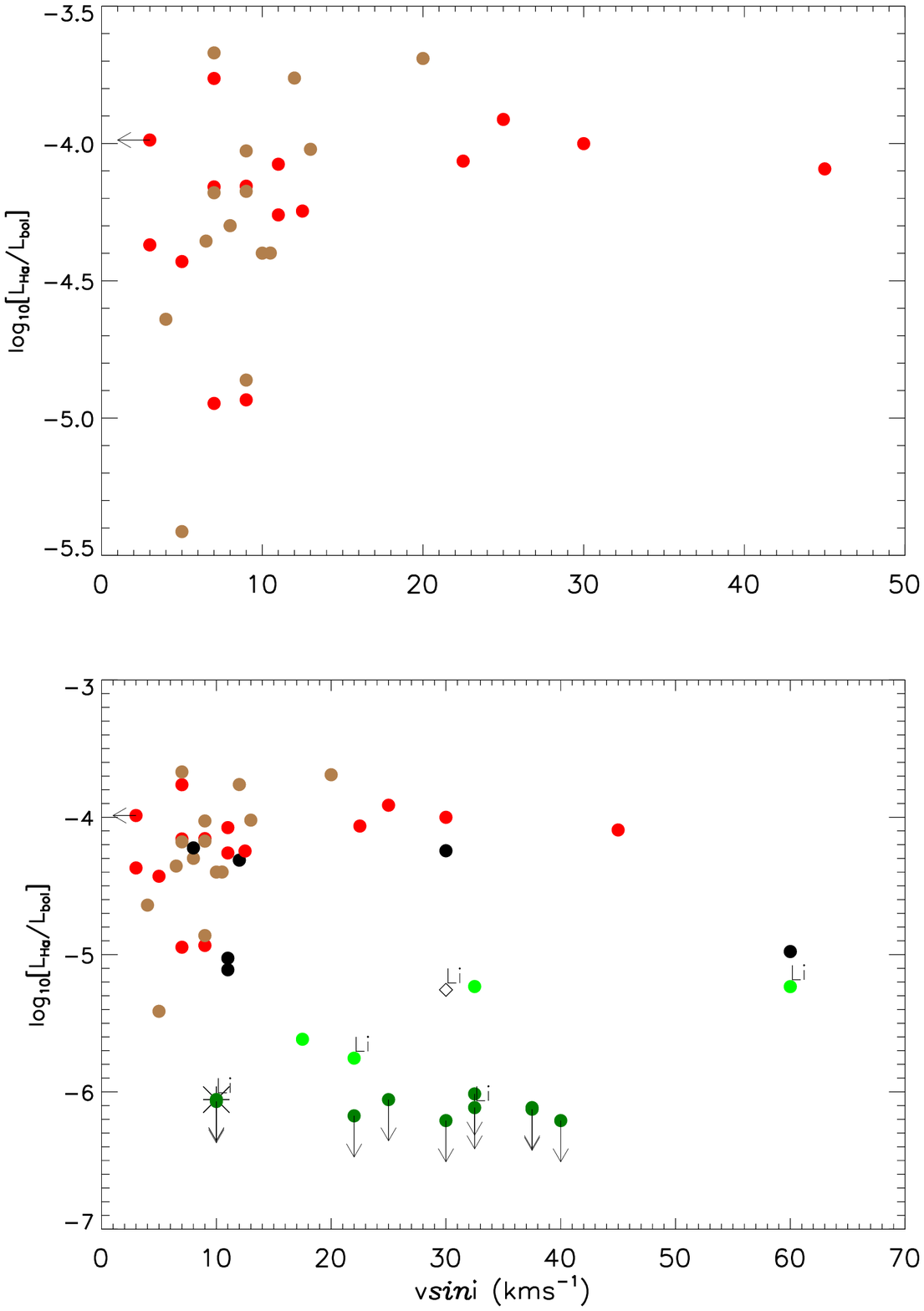}{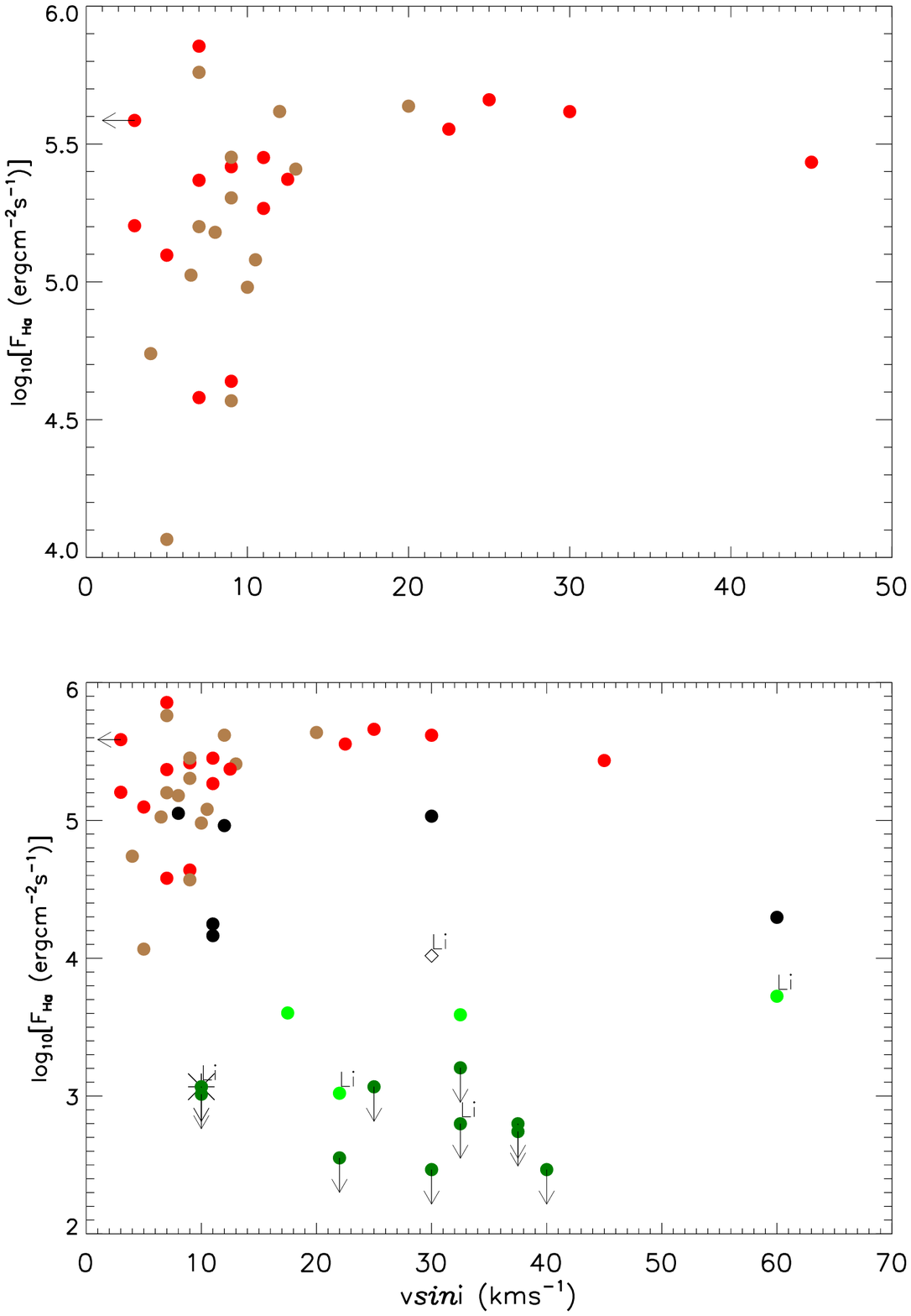}
\figcaption{\label{fig3}LEFT: $\Lhal/\Lbol$ versus \vsini for M5 to L6 dwarfs.  {\it Top Panel}:  M5 - M8.5.  M5-6.5 in red; M7-8.5 in brown.  {\it Bottom panel}:  Our entire sample down to L6.  M9-9.5 in black; L0 and later, with \hal emission, in light green; L0 and later, with no detected \hal emission, in dark green (these objects are also marked with downward arrows).  Diamond marks LP 944-20, a confirmed brown dwarf that has been observed to flare in X-rays and radio.  RIGHT:  $\Fhal$ (\hal surface flux) versus \vsini.  Otherwise same as left panels.  In the M5 - M8.5 dwarfs, using surface flux instead of luminosity ratio as activity indicator seems to give a tighter rotation-activity relation (the two outliers with the highest $\Fhal$ are flare stars), but perhaps not enough to choose one over the other as a more fundamental measure of chromospheric activity.  }

\plotfiddle{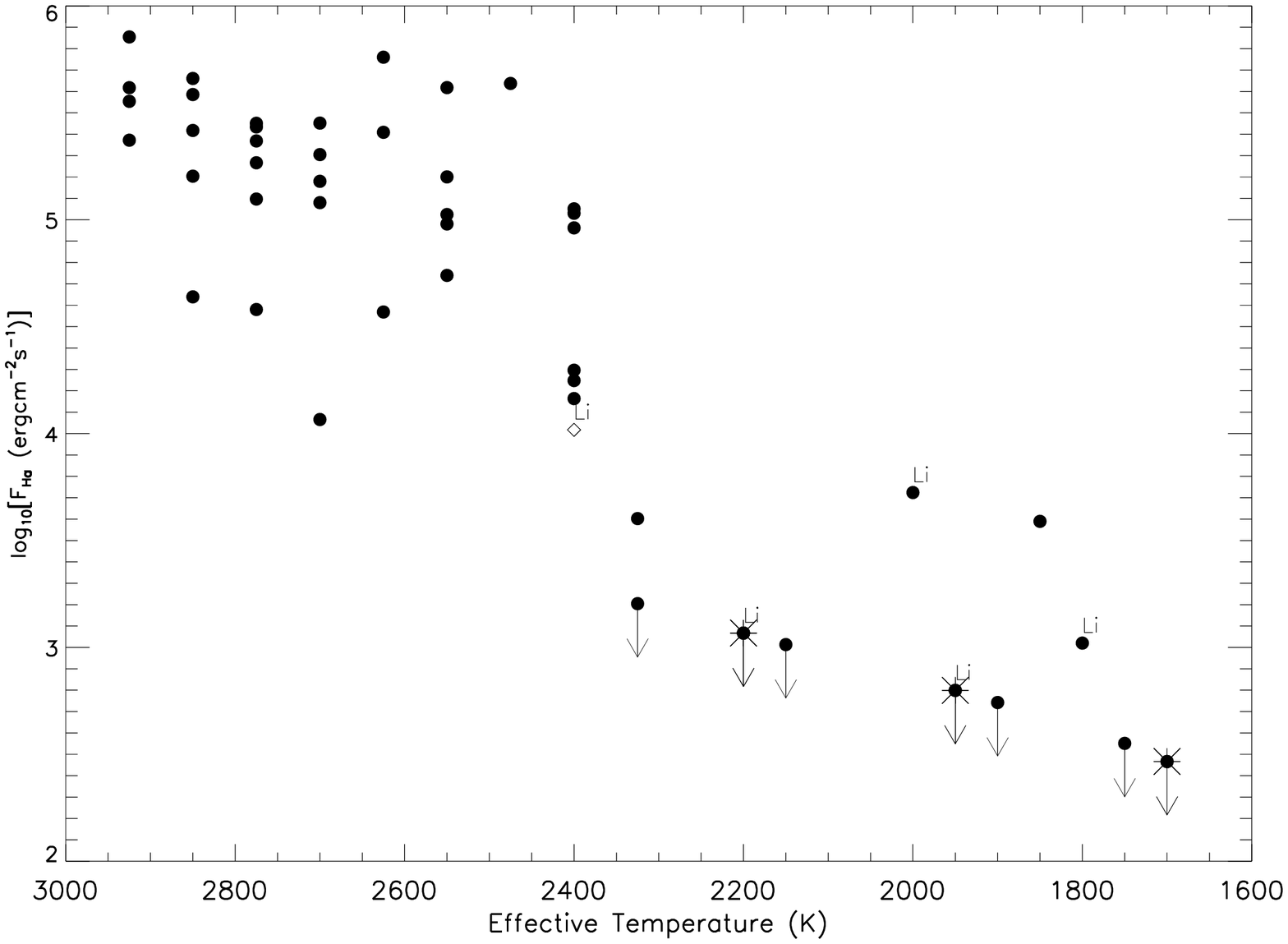}{7cm}{0}{45.}{45.}{-150cm}{0cm}
\figcaption{\label{fig4} $\Fhal$ versus \teff for M5 - L6 dwarfs.  Note rapid falloff in activity at about 2400 K ($\sim$ M9).  Plot reveals that the decline in activity with later type may be a direct result of decreasing \teff.  Note that the slope of the envelope of undetected $\Fhal$, for \teff $\leq$ 2300 K, reflects the improvement in \hal detection limits with decreasing \teff (since the background photosphere becomes fainter), and not a real trend in $\Fhal$; the actual \hal surface flux here could be much lower. }


\begin{thebibliography}{}
\bibitem[Basri 2000]{basri2000}
Basri,G., 2000, {\it Eleventh Cambridge Workshop on Cool Stars, Stellar Systems and the Sun}, eds. Garc\'{i}a-Lop\'{e}z, Rebolo, Zapatero-Osorio
\bibitem[Charbonneau et al., 1997]{charb97} 
Charbonneau,P., Schrijver,C.J., Macgregor, K.B., 1997, {\it Cosmic Winds and the Heliosphere}, eds. Jokipii,J.R., Sonett,C.P., Giampapa,M.S. (The University of Arizona Press)
\bibitem[Delfosse et al., 1998]{delfosse98}
Delfosse,X., Forveille,T., Perrier,C., Mayor,M., 1998, \aap, 331, 581
\bibitem[Durney et al., 1993]{durney93}
Durney,B.R., De Young,B.S., Roxburgh,I.W. 1993, Solar Physics, 145, 207 
\bibitem[Meyer \& Meyer-Hofmeister, 1999]{hof99}
Meyer,F., \& Meyer-Hofmeister,E., 1999, \aap, 341, L23 
\bibitem[Mohanty \& Basri, submitted]{mohantysub}
Mohanty,S., Basri,G., submitted

\end{thebibliography}
\end{document}